\documentclass[twocolumn]{article}
\usepackage{imr}
\usepackage{hyperref}
\usepackage{amsmath}
\usepackage{amssymb}
\usepackage{textcomp}
\usepackage{graphicx}
\usepackage{algorithm2e}
\usepackage{todonotes}
\usepackage{multirow}
 \usepackage{listings}
\usepackage{tikz}

\usepackage{booktabs}
\usepackage{array}
\newcolumntype{L}[1]{>{\raggedright\let\newline\\\arraybackslash\hspace{0pt}}m{#1}}
\newcolumntype{C}[1]{>{\centering\let\newline\\\arraybackslash\hspace{0pt}}m{#1}}
\newcolumntype{R}[1]{>{\raggedleft\let\newline\\\arraybackslash\hspace{0pt}}m{#1}}
\usepackage{subcaption}
\usepackage{pgfplots}

\pgfplotsset{compat=newest}
\pgfplotsset{yticklabel style={text width=2em,align=right}}
\newlength\figureheight
\newlength\figurewidth

\definecolor{UCLorange}{HTML}{EB811B}
\definecolor{UCLdarkBlue}{RGB}{82, 100, 128}
\definecolor{UCLred}{RGB}{255,102,102}
\definecolor{UCLdarkBlue}{RGB}{82, 100, 128}
\definecolor{UCLgreen}{RGB}{80,150,0}
\colorlet{UCLdarkRed}{black!40!UCLred}
\colorlet{UCLdarkGrey}{white!30!black}

\begin{document}

\title{\uppercase{Hex Me If You Can}}
\author{Pierre-Alexandre Beaufort$^1$, Maxence Reberol$^2$, Heng Liu$^1$, Franck Ledoux$^{3,4}$, David Bommes$^1$}
\date{
  $^1$University of Bern, CGG, 3012 Bern, Switzerland\\  
  $^2$Université catholique de Louvain, iMMC, 1348 Louvain-la-Neuve, Belgium\\  
  $^3$CEA, DAM, DIF, F-91297, Arpajon, France\\
  $^4$Université Paris-Saclay, CEA, LIHPC, 91680 Bruyères-le-Châtel, France
}

\abstract{
  HEXME consists of tetrahedral meshes with tagged features, and of a workflow to generate them.
  The main purpose of HEXME meshes is to enable consistent and fair evaluation of hexahedral meshing algorithms and related techniques.
  The tetrahedral meshes have been generated with Gmsh, starting from 63 computer-aided design (CAD) models coming from various databases.
  To highlight and label the various and challenging aspects of hexahedral mesh generation, the CAD models are classified into three
  categories: simple, nasty and industrial. For each CAD model, we provide three kinds of tetrahedral meshes.
  The mesh generation yielding those 189 tetrahedral meshes is defined thanks
  to Snakemake, a modern workflow management system, which allows us to define
  a fully automated, extensible and sustainable workflow.  It is possible to
  download the whole dataset or to pick some meshes by browsing the online
  catalog. Since there is no doubt that the hexahedral meshing techniques are
  going to progress, the HEXME dataset is also built with evolution in mind.
  A public GitHub repository hosts the HEXME workflow, in which
  external contributions and future releases are possible and encouraged.  
}


\maketitle
\thispagestyle{empty}
\pagestyle{empty}

\section{Introduction}

Structured hexahedral meshes are highly desired to perform fast, accurate and adaptive numerical simulations.
For an even number of vertices, there are at least five times less hexahedra than tetrahedra\cite{10.1145/3272127.3275037}.
Hexahedral finite elements offer a richer functional space, and if they are structured, they may provide discrete schemes whose effectiveness is close to finite difference methods.
The regularity\footnote{almost everywhere} of the local topology allows to perform anisotropic refinements where finer approximations are needed.

\begin{figure}[htb]
	\centering
	\includegraphics[width=\linewidth]{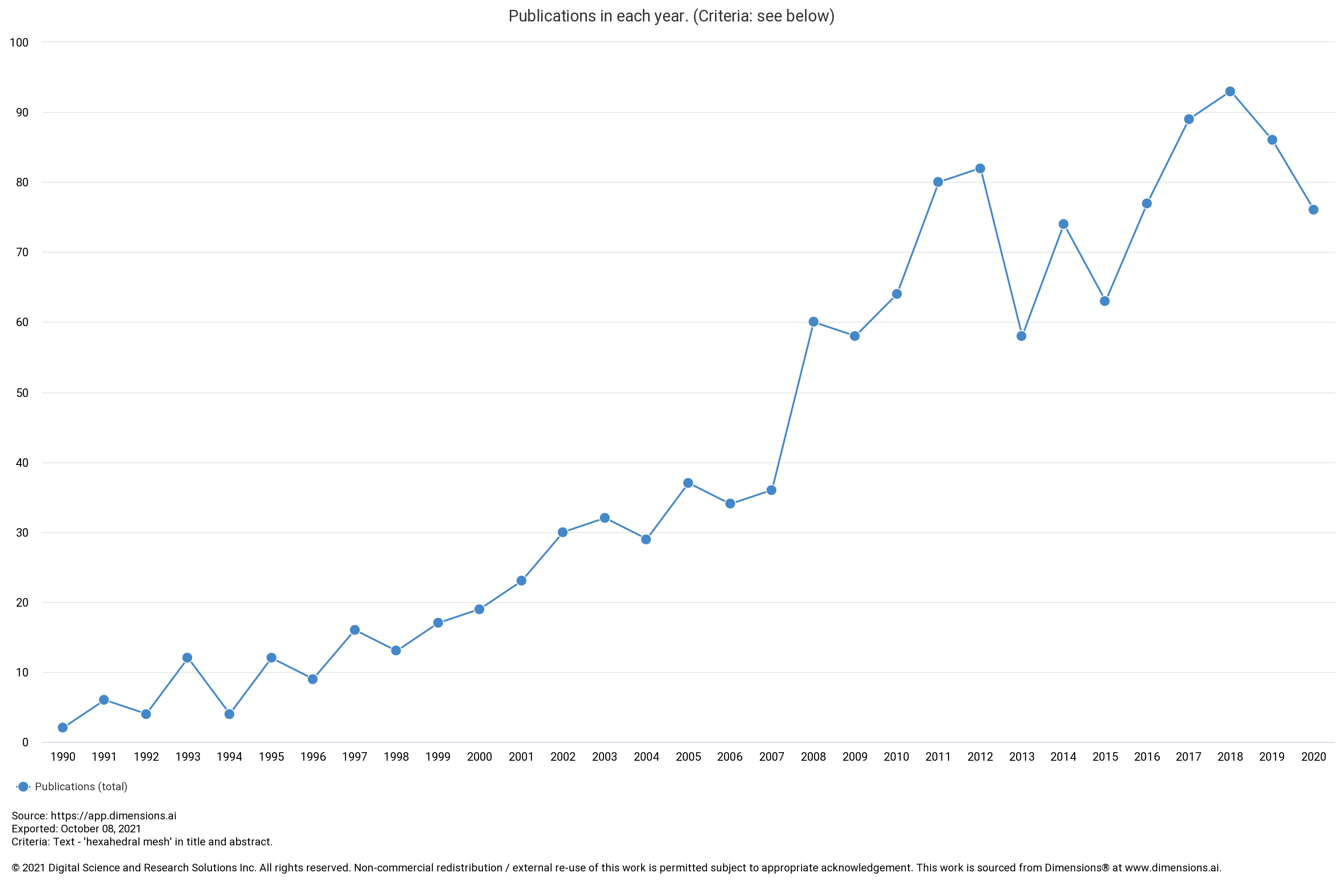}
	\caption{Yearly number of publications related to hexahedral meshes. Source: \href{https://app.dimensions.ai}{app.dimension.ai} .}
	\label{fig:hex-per-year}
\end{figure}

Those last decades, the mesh community has been actively working on developing auxiliary tools, in order to implement a robust and automatic hexahedral mesher, Fig.\ref{fig:hex-per-year}.
Although numerous researches and works have been done, this numerical dream remains an unsolved challenge up to this day.
Indeed, there is no \emph{automatic} method yielding \emph{robustly} a high \emph{quality} hexahedral mesh.
Industrial methods manage to provide high quality meshes, but they typically involve tedious and lengthy user interventions.
Automatic methods such as advanced frontal\cite{baudouin2014frontal} and polycube-based\cite{cherchi2019selective} approaches are not guaranteed to be robust, neither to build high quality hexahedra.
Solely octree-based procedures\cite{gao2019feature} can automatically and robustly supply a full hexahedral mesh, but with a quality far from ideal on the boundary.
Three-dimensional octahedral frame-based methods are able to automatically and robustly make a high quality mesh, \emph{provided} that the octahedral singularities (i.e. not a frame) consistently match irregular configurations (eg. an edge shared by 5 hexahedra)\cite{liu2018singularity} of hexahedral grids. 

Most of the algorithms tackling the hex meshing challenge use tetrahedral meshes in the background.
Yet, there misses a tetrahedral dataset to perform objective analyses and fair comparisons of state-of-the-art hexahedral methods.
Moreover, enabling such a dataset would allow identifying common issues, hence providing awareness of improvement clues.

Here is HEXME dataset, a collection of tetrahedral meshes with tagged feature entities.
The feature entities are special points, curves and/or surfaces, which should be factually captured by a hexahedral mesh.
All meshes have been generated from computer-aided design (CAD) models, following a worflow defined with Snakemake\cite{molder2021sustainable}, using Gmsh\cite{geuzaine2009gmsh} API with custom parameters defined in yaml metadata files.
CAD models are classified in three categories (simple, nasty, industrial), in order to grade their difficulty and consistency.
For each model, three meshes are provided: two resolutions (coarse, uniform) to analyze the mesh dependency, and a playground (bounding) to dare customizable methods.
The meshes are exported as vtk datafiles (version 2, ascii mode), a mesh format which is broadly used.

This paper first describes datasets whose content is comparable with HEXME.
Afterwards, the pipeline having produced the tetrahedral meshes from CAD models is fully exposed.
HEXME datatset is then dessicated to present what is available \emph{as-is}.
{\color{gray}Finally, some hexahedral methods are applied to the dataset, in order to give some insight about HEXME purpose.}

\section{Comparable Datasets}\label{sec:comparable-datasets}

\paragraph{Tetwild} Even though Tetwild\cite{Hu:2018:TMW:3197517.3201353} is a tetrahedral meshing technique, it is also a tetrahedral \href{https://drive.google.com/file/d/17AZwaQaj_nxdCIUpiGFCQ7_khNQxfG4Y/view?usp=sharing}{dataset}, since the authors provide the output of their algorithm applied to Thingi10k\cite{Thingi10K}, a triangular dataset. This tetrahedral dataset is the tetrahedrization of ten thousand models from \href{https://ten-thousand-models.appspot.com/}{Thingi10k}. The tetrahedral meshes are \texttt{msh2} binary files, with a scalar field per tetrahedron corresponding to their respective minimal diheral angle. The 10k meshes are stored on Google drive, within an archive \texttt{tar.gz}.

\paragraph{ABC} The ABC\cite{Koch_2019_CVPR} \href{https://deep-geometry.github.io/abc-dataset/}{dataset} is a collection of one million computer-aided design models for geometric deep learning. All CAD files are from \href{https://www.onshape.com/en/}{Onshape}, and the original information related to those models is recorded within a metadata file \emph{meta}\texttt{.yml}. Some processing tasks are done in order to filter the duplicate and broken models. The filtered models \emph{para}\texttt{.zip} are afterwards converted into \emph{step}\texttt{.step} and \emph{stl2}\texttt{.stl} files, using \href{https://www.plm.automation.siemens.com/global/en/products/plm-components/parasolid.html}{Parasolid}. Gmsh\cite{geuzaine2009gmsh} is then used to provide higher quality triangular meshes (either uniform, or adapted to the curvature), which are exported as \emph{obj}\texttt{.obj} meshes, from the \texttt{.step} files. Differential quantities are stored in those  \emph{obj}  files, while the vertices and triangles of the mesh are respectively matched to the feature curves and patches, through another metadata file \emph{feat}\texttt{.yml}. Further files may be provided, depending on the success of the processing. The dataset is downloadable by chunks containing \texttt{7z} archives of above files.

\paragraph{Thingi10k} Historically, Thingi10k\cite{Thingi10K} is the first \href{https://ten-thousand-models.appspot.com/}{dataset} providing ten thousand various, complex and quality \texttt{.stl} triangulations of 3D (printing) models. All models come from \href{https://www.thingiverse.com/}{Thingiverse}, and have been selected only if they are tagged \emph{featured} by the Thingiverse staff. An online query interface is provided, which returns all the contextual and original information related to a stl triangulation. This interface is the only way to access to the dataset.

\paragraph{SimJEB} A recent\footnote{2021} \href{https://simjeb.github.io/}{dataset}, SimJEB\cite{10.1111:cgf.14353}, provides 381 tetrahedral meshes from CAD models, by following a \emph{semi-automated} pipeline. The CAD models come from a \href{https://grabcad.com/challenges/ge-jet-engine-bracket-challenge}{challenge} organized by GrabCAD. Those former 700 models have been filtered (mostly based on the filename), manually repaired, and then meshed using the commercial software \href{https://www.altair.com/hypermesh/}{HyperMesh}. Eventually, a structural simulation was performed using the commercial software \href{https://www.altair.com/optistruct/}{OptiStruct}. The 381 \texttt{.vtk} tetrahedral meshes surviving this pipeline are hosted through the Harvard Dataverse, along with the corresponding clean CAD \texttt{.stp} file, triangular surface \texttt{.obj} meshes, finite element \texttt{.fem} models, and simulation \texttt{.csv} results. The final models are identified by an integer, whose origin is tracked by some \emph{readme} files. A web page allows to browse the designs, and to explore the data.

\section{From CAD to Tets}\label{sec:cad2tets}

\begin{figure*}[!htbp]
  \centering
  \captionsetup{justification=centering}
  \begin{minipage}[c]{.2\textwidth}
	\begin{subfigure}[t]{\textwidth}
	    \centering
	    \includegraphics[width=\textwidth]{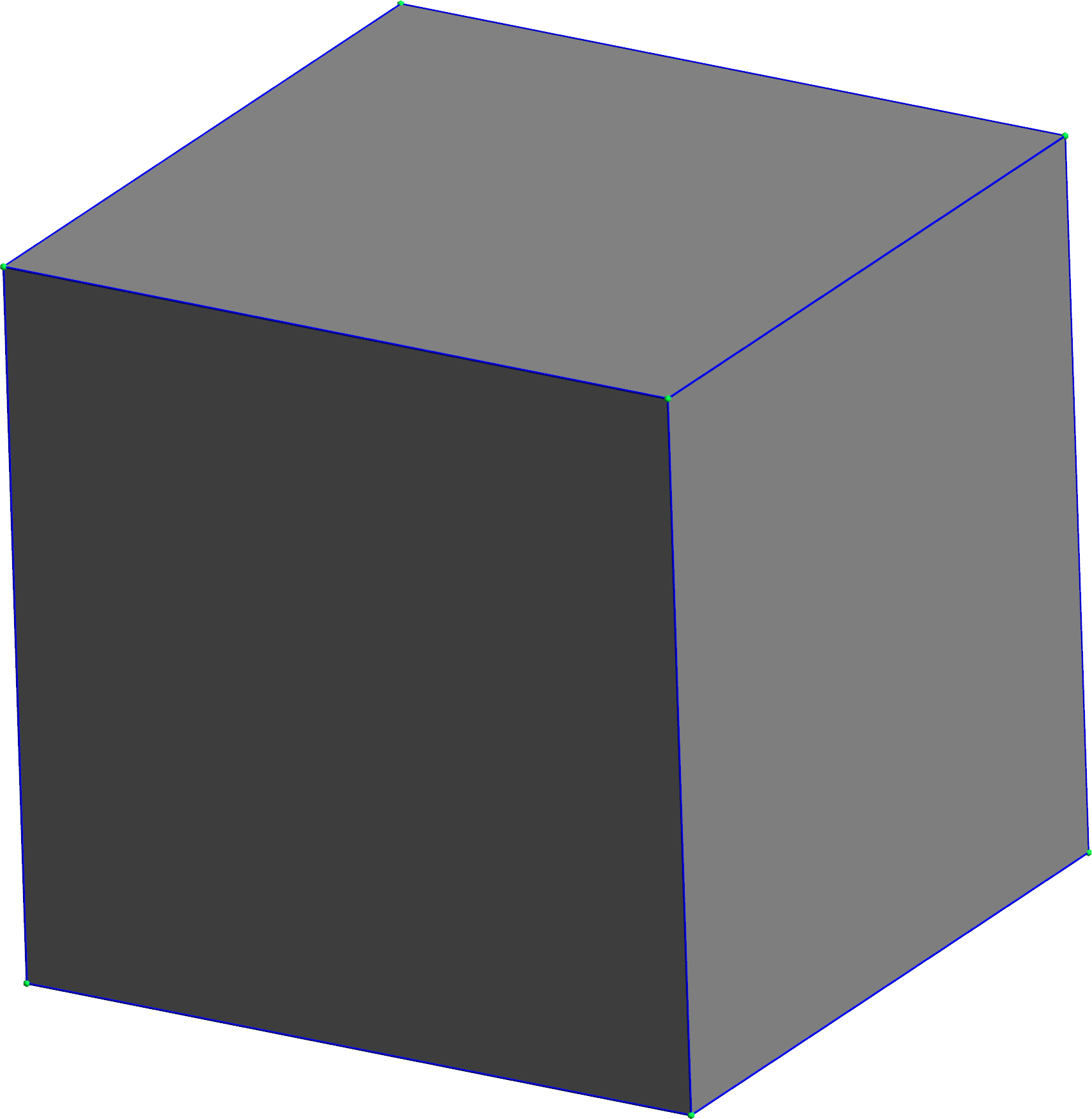}            
            \caption{A simple model: \texttt{{\color{UCLdarkRed}s}01o\_cube.geo}}
            \label{sub-fig:simple-cad}
	\end{subfigure}
        \begin{subfigure}[t]{\textwidth}
	    \centering
	    \includegraphics[width=\textwidth]{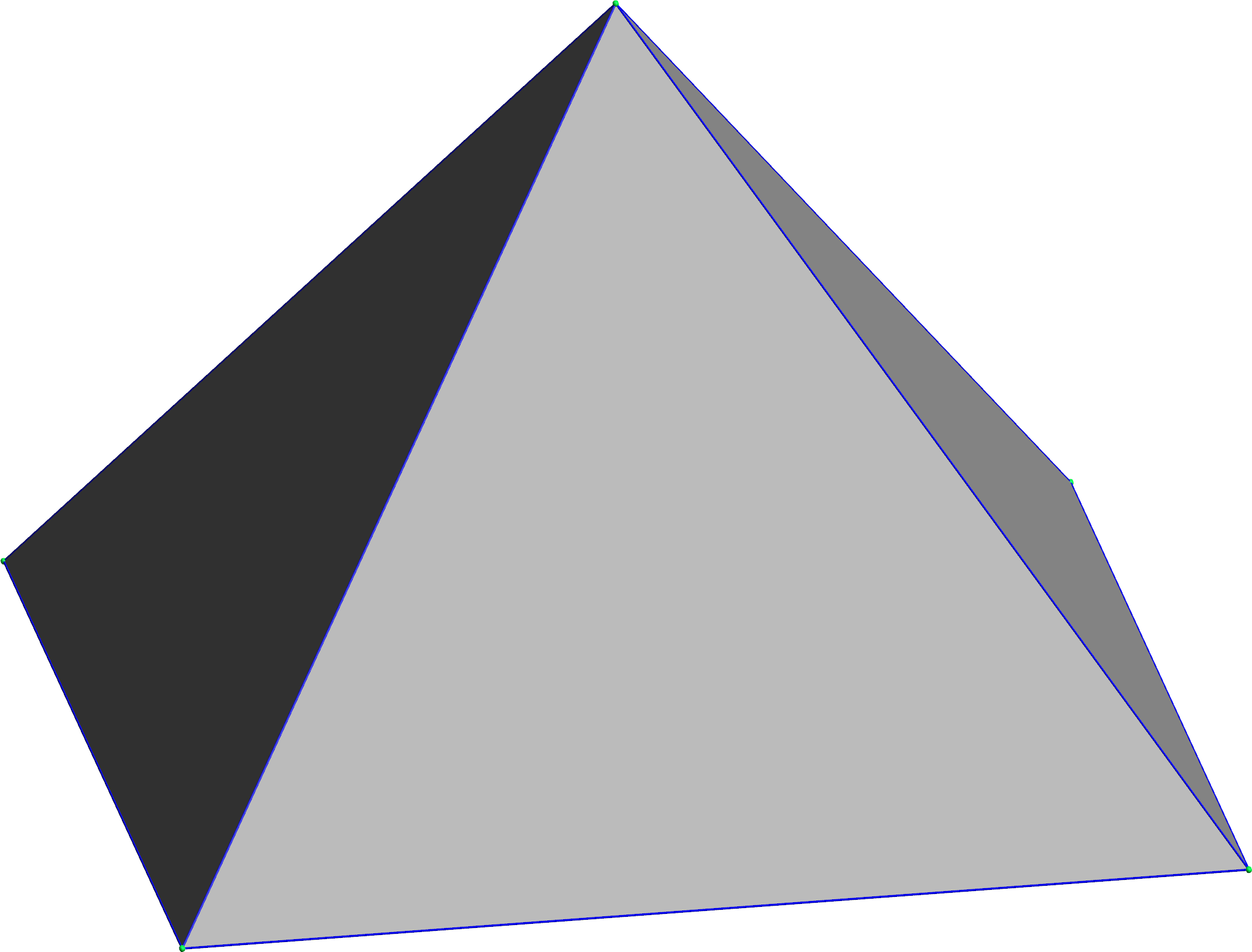}
            \caption{A nasty model: \texttt{{\color{UCLgreen}n}09o\_pyramid.geo}}
            \label{sub-fig:nasty-cad}
	\end{subfigure}
  \end{minipage}
  \begin{minipage}[c]{.7\textwidth}
        \begin{subfigure}[t]{\textwidth}
	    \centering
	    \includegraphics[width=\textwidth]{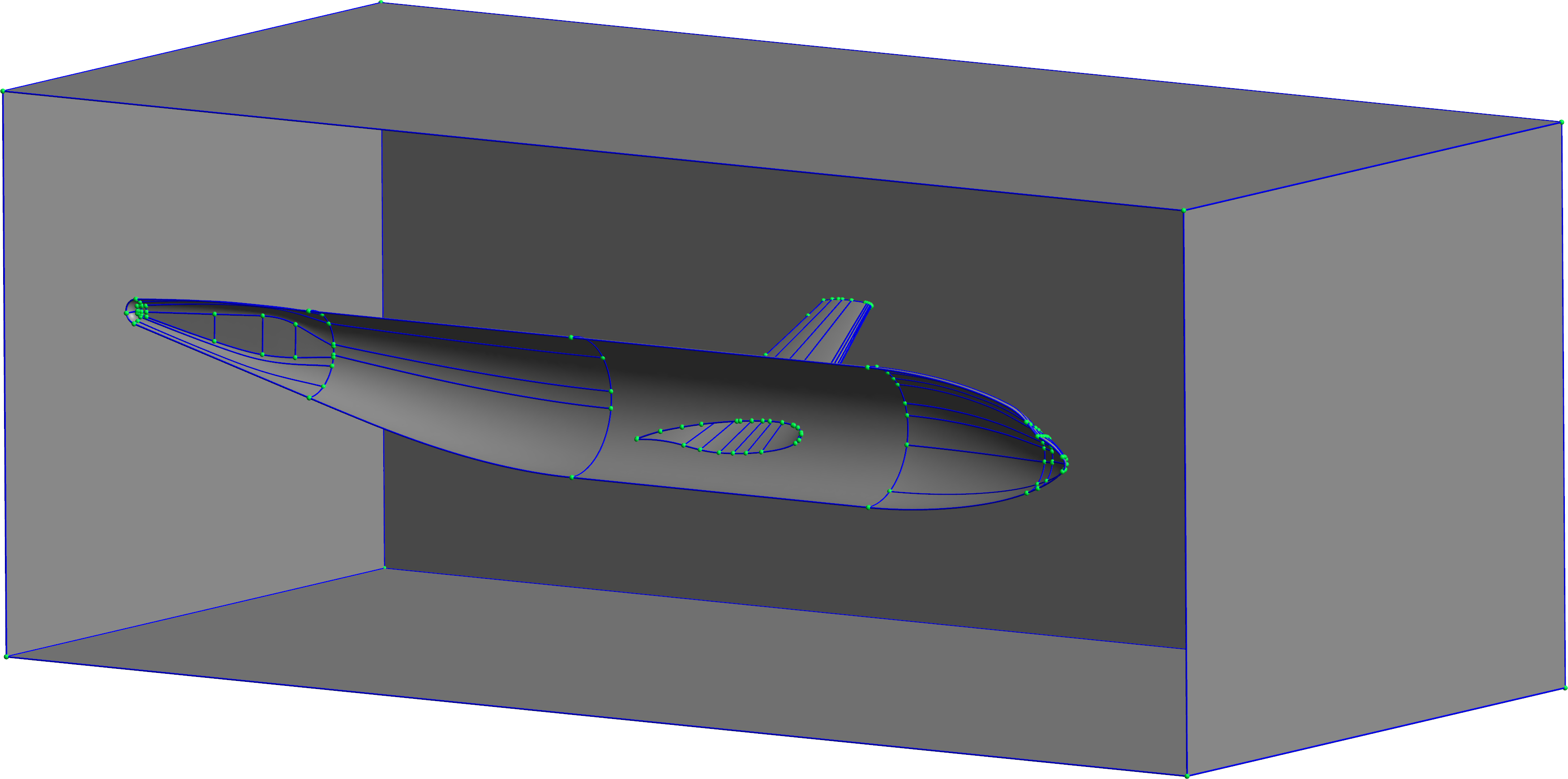}
            \caption{An industrial model: \texttt{{\color{blue}i}31o\_dlr\_f6.brep}\cite{vassberg2008abridged, brodersen2008dlr}} (the front box face is not shown) 
            \label{sub-fig:industrial-cad}
	\end{subfigure}
  \end{minipage}
	\caption{HEXME uses three categories of CAD models.}
	\label{fig:cad-examples}
\end{figure*}

All the tetrahedral meshes provided by HEXME have been produced from three categories of CAD models:
\begin{itemize}
\item \textbf{{\color{UCLdarkRed}s}}imple models: basic shapes that are assumed to be easily hex-meshable, i.e. the target hexahedral topology is fair (eg. a cube, Fig.\ref{sub-fig:simple-cad}).
\item \textbf{{\color{UCLgreen}n}}asty models: academic shapes that are challenging to hex-mesh (eg. a pyramid\cite{VERHETSEL2019102735}, Fig.\ref{sub-fig:nasty-cad}).
\item \textbf{{\color{blue}i}}ndustrial models: lifelike shapes whose hexahedrization is highly valuable for numerical simulation (eg. an aircraft model for CFD, Fig.\ref{sub-fig:industrial-cad}).
\end{itemize}

The 63 CAD models have been arbitrarily selected.
Nevertheless, we believe that those models already represent most of the situations/challenges that hexahedral meshers face at the moment.
Besides, their number is low enough to allow human assessment of hexahedral methods from this dataset.

Instead of the comparable datasets (§\ref{sec:comparable-datasets}), the CAD models come from several databases: \href{https://deep-geometry.github.io/abc-dataset/}{ABC dataset} (originally from \href{https://www.onshape.com/en/}{Onshape}), \href{https://grabcad.com/}{GrabCAD} and \href{https://gitlab.com/franck.ledoux/mambo}{MAMBO}.
Some CAD models have been created by us, using \href{https://gmsh.info/}{Gmsh} and \href{https://www.plm.automation.siemens.com/global/en/products/nx/}{Siemens NX} softwares. Those latter models are in the \emph{public domain}, while the other ones are regulated by licenses, which are respectively: \href{https://www.onshape.com/en/legal/terms-of-use#your_content}{Onshape Terms 1(g)(i)}, \href{https://grabcad.com/terms}{GrabCAD Terms} (\href{https://help.grabcad.com/article/246-how-can-models-be-used-and-shared}{cf. related FAQ}) and \href{https://www.apache.org/licenses/LICENSE-2.0.html}{Apache 2.0}.
The above information is reported within a metadata file \texttt{(s|n|i)(\textbackslash d\{2\})o\_\{extra\}.yaml} per CAD model, with a short description of the shape (eg.listing \ref{lst:i28o}).

\begingroup
\tiny
\begin{lstlisting}[caption={i28o\_gc\_tire\_1218.yaml}, label={lst:i28o}]
author: Milos Suvakov, https://grabcad.com/milos.suvakov
description: tire of a truck
license: GrabCAD Terms
name: i28o_gc_tire_1218
original: original/i28o_gc_tire_1218.step
references: https://grabcad.com/library/tire-12-00-18-1
\end{lstlisting}
\endgroup

\begin{figure*}[!htbp]
  \centering
  \begin{minipage}[c]{.3\textwidth}
	\begin{subfigure}[t]{\textwidth}
	    \centering
	    \includegraphics[width=\textwidth]{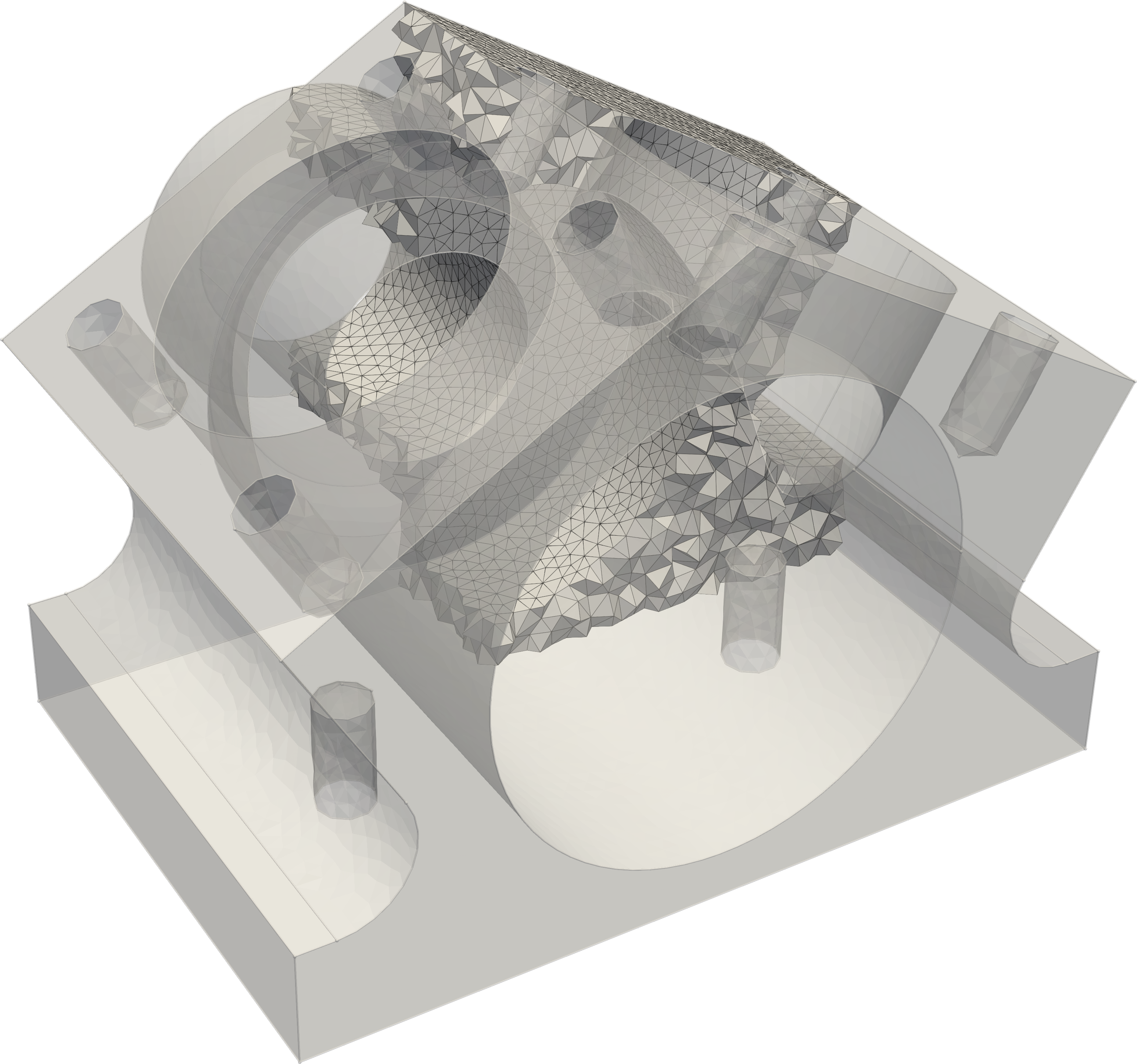}            
            \caption{Coarse model: \texttt{i05{\color{UCLdarkRed}c}\_m5.vtk}}
            \label{sub-fig:coarse-mesh}
	\end{subfigure}
        \begin{subfigure}[t]{\textwidth}
	    \centering
	    \includegraphics[width=\textwidth]{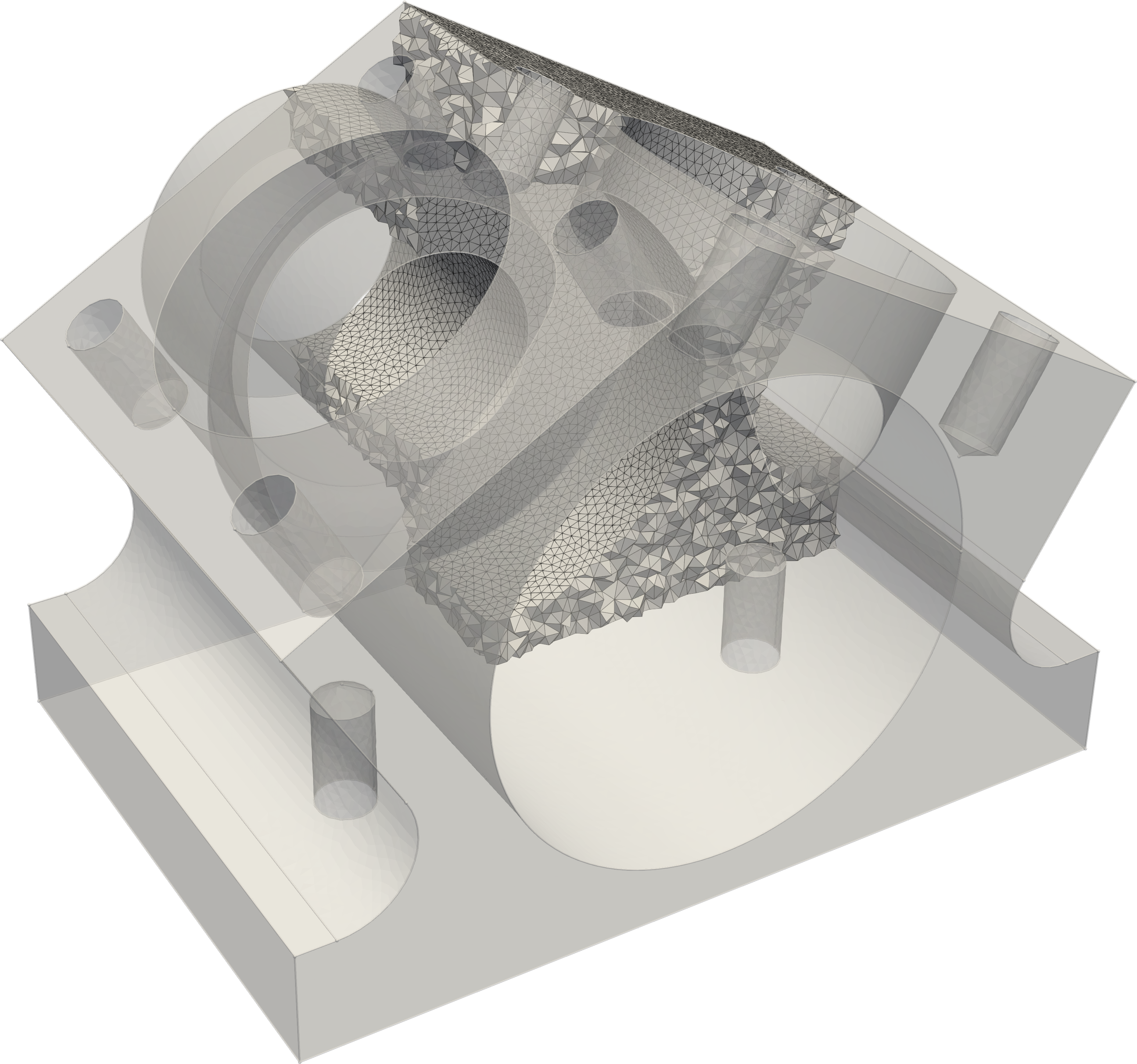}
            \caption{Uniform mesh: \texttt{i05{\color{UCLgreen}u}\_m5.vtk}}
            \label{sub-fig:uniform-mesh}
	\end{subfigure}
  \end{minipage}
  \begin{minipage}[c]{.6\textwidth}
        \begin{subfigure}[t]{\textwidth}
	    \centering
	    \includegraphics[width=\textwidth]{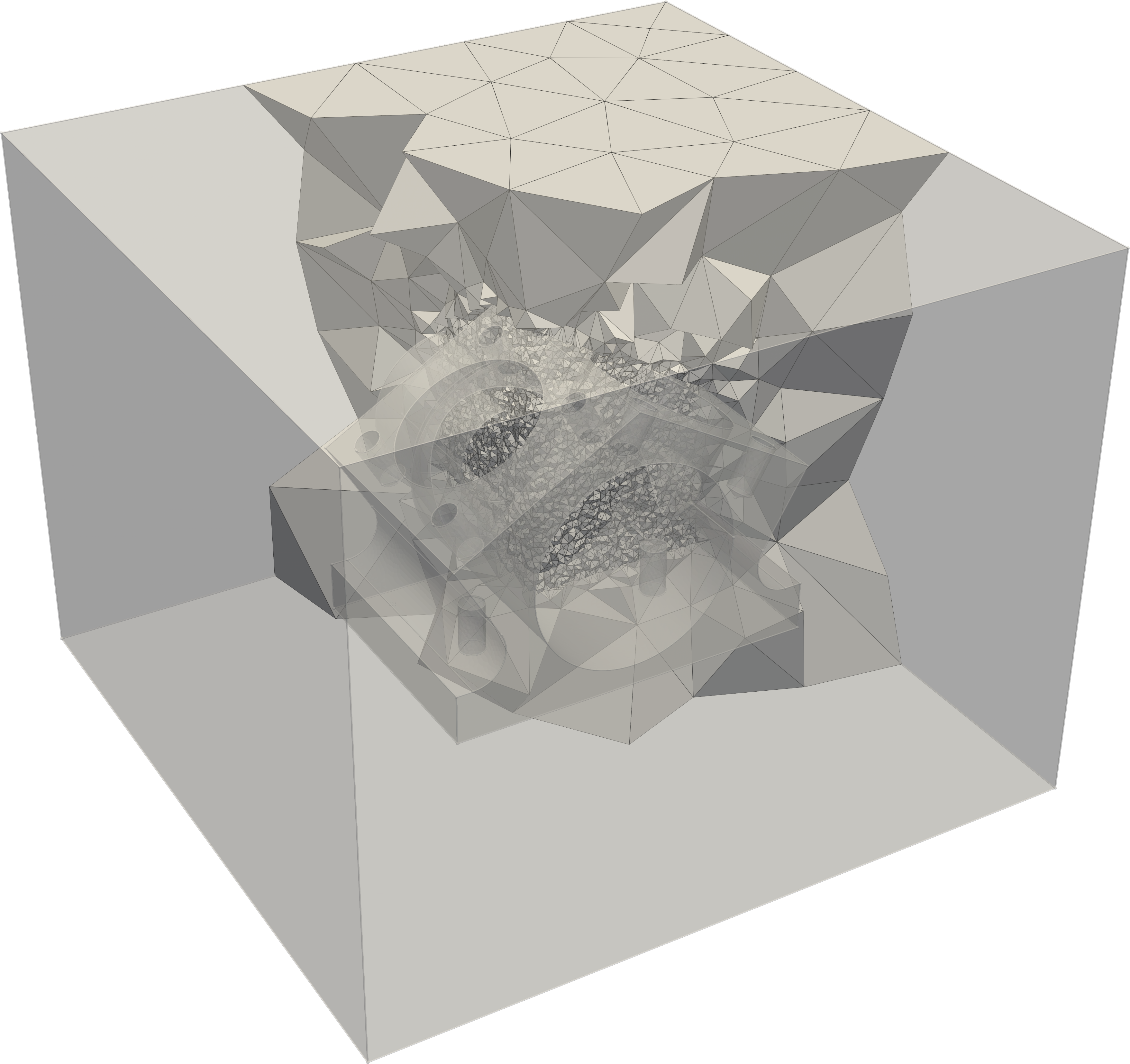}
            \caption{Bounding mesh: \texttt{i05{\color{blue}b}\_m5.vtk}}
            \label{sub-fig:bounding-mesh}
	\end{subfigure}
  \end{minipage}
	\caption{There are three tetrahedral meshes per CAD model, eg. \texttt{i05o\_m5.step} from MAMBO.}
	\label{fig:mesh-examples}
\end{figure*}

For every CAD model, three tetrahedral meshes are provided:
\begin{itemize}
\item \textbf{\color{UCLdarkRed}c}oarse mesh: the mesh size is only upper bounded s.t. the geometry is maintained (eg. Fig.\ref{sub-fig:coarse-mesh}).
\item \textbf{\color{UCLgreen}u}niform mesh: the mesh size is constant, even in the neighborhood of the tiniest geometrical features (eg. Fig.\ref{sub-fig:uniform-mesh}).
\item \textbf{\color{blue}b}ounding mesh: the initial model is immersed in a box that is as twice as large as the original bounding box, and the corresponding mesh is such that the smallest gap between the box side and the inital model is meshed by one layer of tetrahedra (eg. Fig.\ref{sub-fig:bounding-mesh}).
\end{itemize}

The pipeline handling the mesh generation is orchestrated by \href{https://snakemake.github.io/}{Snakemake}\cite{molder2021sustainable}, a \href{https://badge.dimensions.ai/details/id/pub.1018944052}{popular}\footnote{$\sim 5$ citations per week} scalable workflow management system.
In few words, Snakemake is a modern version of Makefile, whose syntax is close to Python.
The workflow \href{https://github.com/cgg-bern/hex-me-if-you-can/blob/main/Snakemesh}{\texttt{Snakemesh}} consists of two rules.
On one hand, the first rule \texttt{meshes} defines which meshes should be produced.
On the other hand, the second rule \texttt{cad2vtk} generates a mesh from CAD model and a metadata file \texttt{(s|n|i)(\textbackslash d\{2\})(\underline{c}|\underline{u}|\underline{b})\_\{extra\}.yaml} containing the custom mesh options (among those: either \textbf{c}oarse, \textbf{u}niform xor \textbf{b}ounding).
To do so, this second rule runs a python script using Gmsh api, with a maximum of 8 threads.
For each mesh, a log file \texttt{(s|n|i)(\textbackslash d\{2\})(c|b|u)\_\{extra\}.\underline{txt}} is written with the corresponding console output, in order to record the history of the meshing task.

Snakemake scans the workflow in a backward fashion, meaning that the input files are identified from the output ones.
In other words, the purpose of the first rule is to state all meshes that should be produced.
Afterwards, the second rule provides those meshes by identifying accordingly the corresponding input files, which are the CAD model and metadata file.
This backward identification is the key of the workflow definition, since the rules are mostly written with wildcards.
The use of Snakemake easily yields a sustainable dataset, since a rule is applied only if an output is older than the corresponding input or missing.

Gmsh\cite{geuzaine2009gmsh} does not mesh only the volume, but also the feature entities as defined by the CAD model.
In addition to the tetrahedral cells, there are triangular, edge and vertex cells\footnote{those lower dimensional cells are conforming to the higher ones} to respectively discretize feature surfaces, curves and points.
Those features are identified by the CAD with a color (i.e. a positive integer), which corresponds to a physical group within Gmsh.
Doing so, the analogous mesh cells are created accordingly with the corresponding CAD color.
Meshes are exported as \href{https://kitware.github.io/vtk-examples/site/VTKFileFormats/}{vtk Datafile Version 2.0}, in ASCII mode.
The used Gmsh git-version is written within the file header.
A mesh is defined as an \texttt{UNSTRUCTURED\_GRID}, with the four following sections:
\begin{enumerate}
\item \texttt{POINTS}: coordinates of every node
\item \texttt{CELLS}: number of nodes, and nodal definition of every element (vertices, edges, triangles and tetrahedra)
\item \texttt{CELL\_TYPES}: integer corresponding to the element type (\{1:vertex, 3:edge, 5:triangle, 10:tetrahedron\})
\item \texttt{CELL\_DATA}: integer corresponding to the element color that belong to the CAD feature.
\end{enumerate}

\begingroup
\tiny
\begin{verbatim}
# vtk DataFile Version 2.0
, Created by Gmsh 4.9.0-git-b39c72341 
ASCII
DATASET UNSTRUCTURED_GRID
POINTS 9332 double
47.96187233897071 17.79329564613556 3.334023045560881e-14
[...]
CELLS 51854 233863
1 0
[...]
CELL_TYPES 51854
1
[...]
CELL_DATA 51854
SCALARS CellEntityIds int 1
LOOKUP_TABLE default
1
[...]
\end{verbatim}
\endgroup

There are then 189 meshes, whose the given filenames \texttt{(s|n|i)(\textbackslash d\{2\})(c|u|b)\_\{extra\}.vtk} summarize the corresponding model \texttt{(s|n|i)}(\textbackslash d\{2\}) and mesh \texttt{(c|u|b)} types.

\section{HEXME Anatomy}

\begin{figure}[htb]
	\centering
	\includegraphics[width=\linewidth]{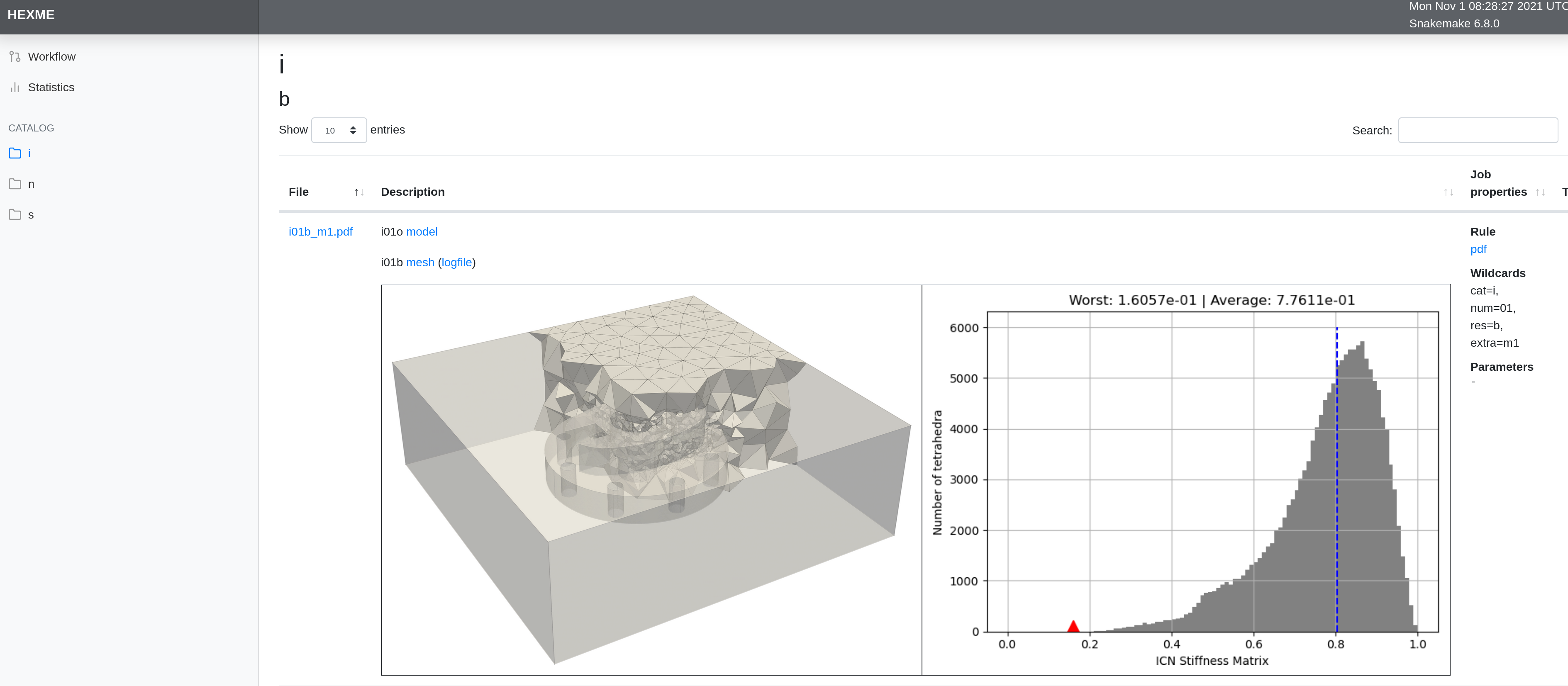}
	\caption{An entry of HEXME catalog: \texttt{i01b\_m1}.}
	\label{fig:hexme-catalog}
\end{figure}

HEXME tetrahedral dataset is downloadable in a single file: \href{https://cgg.unibe.ch/hexme/hexme.zip}{hexme.zip}($\sim$1.5GB, stored on cgg.unibe server).
Otherwise, it is possible to download meshes \emph{one-by-one} from the \href{https://cgg.unibe.ch/hexme}{catalog}\footnote{The catalog is mostly generated by Snakemake, by using the \href{https://snakemake.readthedocs.io/en/stable/snakefiles/reporting.html}{report} feature.} (stored on cgg.unibe server).
The catalog is split into three categories (i, n, s), which corresponds to the model categories (respectively: \textbf{i}ndustrial, \textbf{n}asty, \textbf{s}imple), Fig.\ref{fig:hexme-catalog}.
Within each category, there are three subcategories (b, c, u), which corresponds to the mesh types (respectively: \textbf{b}ounding, \textbf{c}oarse, \textbf{u}niform).

\begin{figure}[htb]
	\centering
	\includegraphics[width=\linewidth]{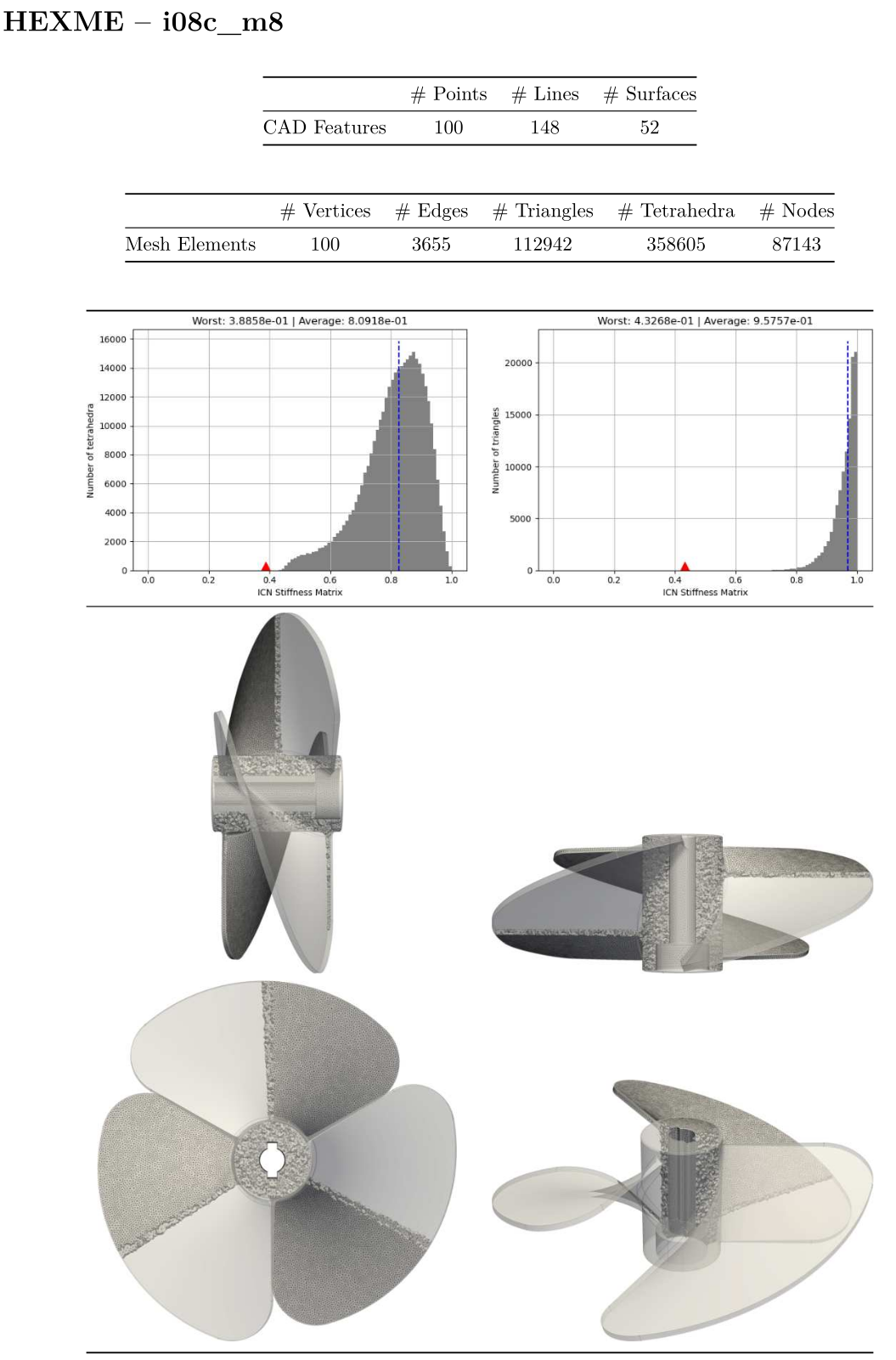}
	\caption{Sheet summarizing \texttt{i08c\_m8}.}
	\label{fig:pdf-example}
\end{figure}

An entry of the catalog is described by two pictures (a cut view and a quality histogram), a \texttt{.pdf} file, a \texttt{.vtk} mesh, the corresponding log file \texttt{(s|n|i)(\textbackslash d\{2\})(c|u|b)\_\{extra\}.txt} and the metadata file \texttt{(s|n|i)(\textbackslash d\{2\})o\_\{extra\}.yaml} related to the CAD model.
The \texttt{.pdf} sheet of a mesh summarizes the mesh, Fig.\ref{fig:pdf-example}.
The summary provides topological information about the CAD model (number of points, curves and surfaces), and the mesh (number of vertices, edges, triangles, tetrahedra and nodes).
Moreover, two histograms related to the inverse condition number (ICN)\cite[\S2.1]{JOHNEN2016328} of triangles and tetrahedra are plotted.
Finally, four screenshots (xy-, yz-, zx- and 3D-views) of the cut mesh are displayed.

\begin{figure}[htb]
  \centering
  \begin{subfigure}[t]{\linewidth}
    \centering
    \includegraphics[width=\linewidth]{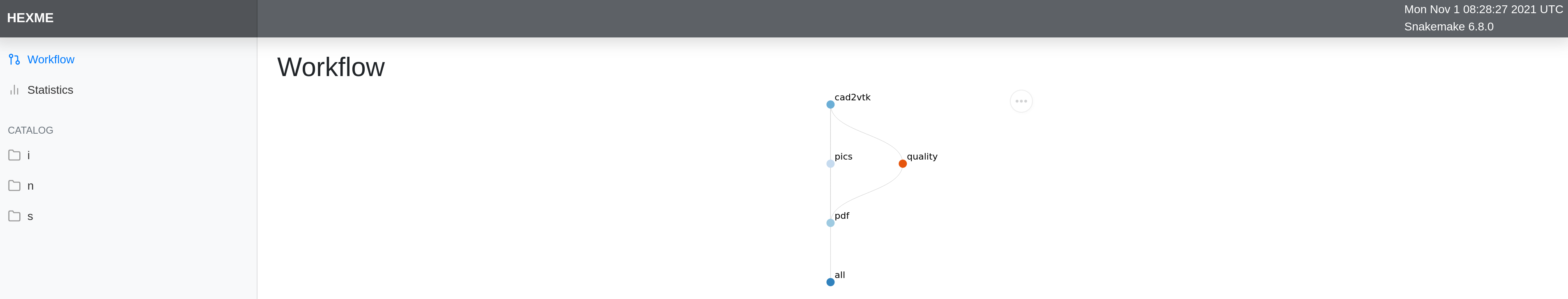}            
    \caption{The whole worflow: from CAD to pdf}
    \label{sub-fig:cad2pdf}
  \end{subfigure}
  \begin{subfigure}[t]{\linewidth}
    \centering
    \includegraphics[width=\linewidth]{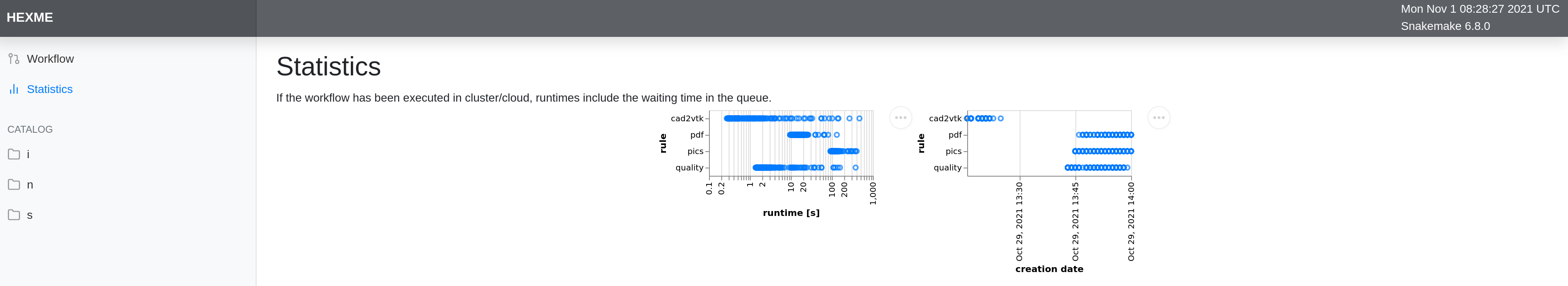}
    \caption{Duration and timestamp}
    \label{sub-fig:times}
  \end{subfigure}
  \begin{subfigure}[t]{\linewidth}
    \centering
    \includegraphics[width=\linewidth]{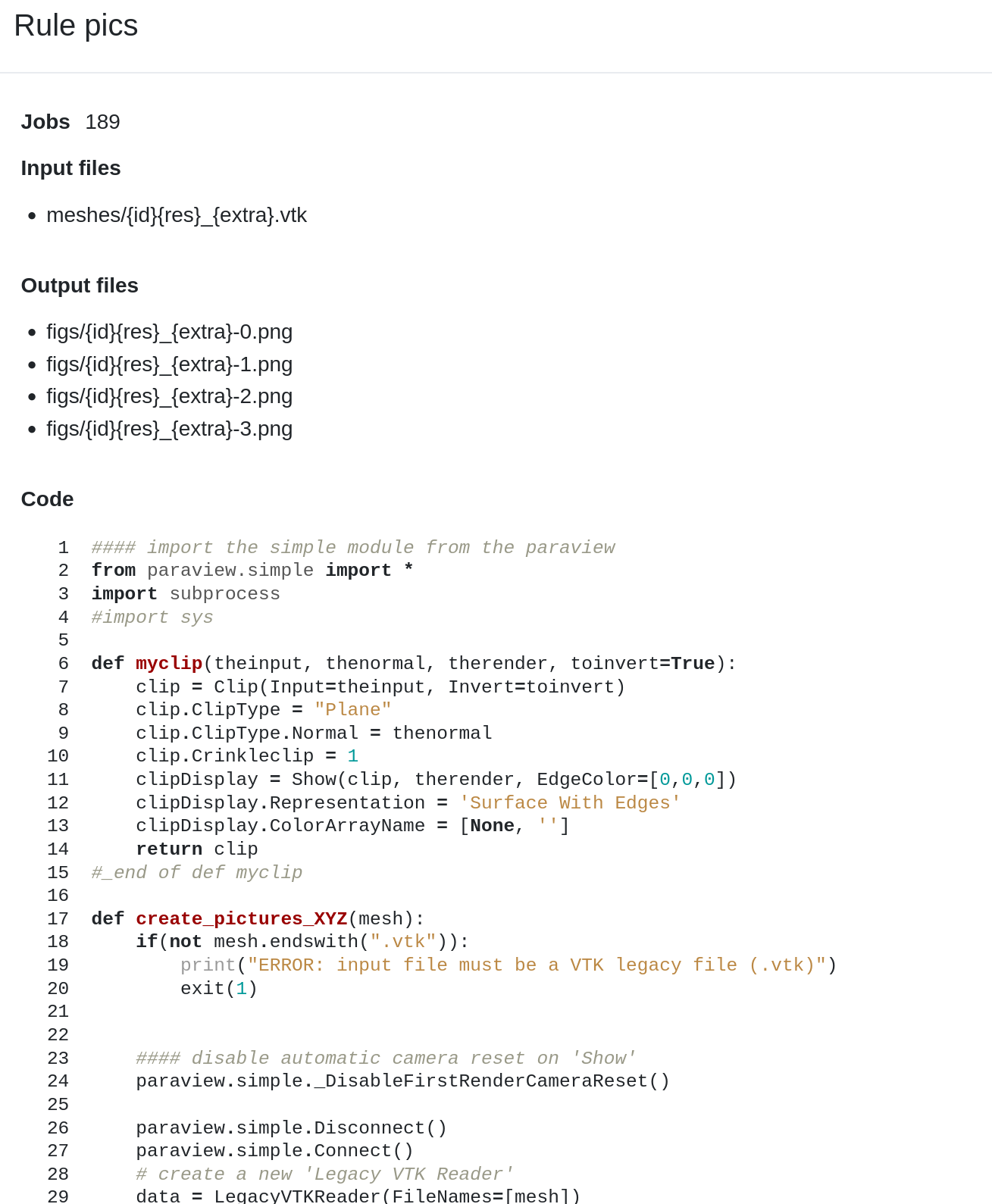}
    \caption{Details about the rule \texttt{pics}}
    \label{sub-fig:rule}
  \end{subfigure}
  \caption{Data about HEXME workflow.}
  \label{fig:cad-examples}
\end{figure}

In addition to those entries, some data related to the workflow are available.
The graph corresponding to the workflow is displayed, Fig.\ref{sub-fig:cad2pdf}.
This graph is longer than what have been introduced into the former section \ref{sec:cad2tets}.
Actually, the views, histograms and sheets are also generated by the workflow.
It is possible to obtain details of a rule by clicking on the corresponding node, Fig.\ref{sub-fig:rule}.
The creation and duration times are also given in statistics, Fig.\ref{sub-fig:times}.

On top of HEXME catalog, there is a public \href{https://github.com/cgg-bern/hex-me-if-you-can}{GitHub} page  hosting all the necessary input files to run the workflow.
The tetrahedral meshes are not hosted on this git repository\footnote{The git history would be too heavy otherwise.}.
The main purpose of this git repository is to expose the workflow that has been used for the mesh generation.
From this repository, it is possible to create \texttt{issue} if one mesh does not meet user expectations.
The mesh community is also invited to contribute to HEXME dataset, by creating \texttt{pull-request} which would consist of proposing new models or/and filtering former ones.
There shall be releases, with appropriate \texttt{git-tag}, whenever the dataset has been significantly updated.

Concretely, HEXME GitHub repository contains:
\begin{itemize}
\item CAD models: \texttt{original/*}
\item metadata files related to the CAD models: \texttt{meta/*o\_*.yaml}
\item metadata files related to the mesh parameters: \texttt{meta/*(b|c|u)\_*.yaml}
\item python scripts involved in the mesh generation and catalog constructions: \texttt{scripts/*.py}
\item the template for an entry of the catalog: \texttt{report/description.rst}
\item the Snakemake files defining the workflow: \texttt{Snakemesh}, \texttt{Snakefile} (\& \texttt{utilies.smk})
\item the virtual machine used to run the mesh generation: \texttt{Dockerfile}
\item the pipeline definition to publish the meshes: \texttt{.gitlab-ci} and \texttt{CI/*}
\item the MIT License for the meshes, and the workflow: \texttt{LICENSE}
\item a \texttt{README.md} file
\end{itemize}
The workflow supports the CAD models whose extension is one of the following: \texttt{.geo}, \texttt{.step}, \texttt{.stp}, \texttt{.brep}.
For every CAD model, four metadata files have to be defined.
The first one gives information about the model
\begingroup
\tiny
\begin{verbatim}
description: ...
license: ...
name: (i|n|s){\d{2}}o_{extra}
original: original/(i|n|s){\d{2}}o_{extra}.(geo|step|stp|brep)
references: ...
\end{verbatim}
\endgroup
The three other ones define the mesh parameters for a mesh type \texttt{(b|c|u)}
\begingroup
\tiny
\begin{lstlisting}[caption={Coarse mesh type}]
gmsh.option.setNumber:
- Mesh.Algorithm: ...
- Mesh.Algorithm3D: ...
- General.NumThreads: ...
- Mesh.CharacteristicLengthMax: ...
- Mesh.MeshSizeFromCurvature: ...
info: meta/(i|n|s){\d{2}}o_{extra}.yaml
\end{lstlisting}
\begin{lstlisting}[caption={Uniform mesh type}]
gmsh.option.setNumber:
- Mesh.Algorithm: ...
- Mesh.Algorithm3D: ...
- General.NumThreads: ...
- Mesh.CharacteristicLengthMin: ...
- Mesh.CharacteristicLengthMax: ...
info: meta/(i|n|s){\d{2}}o_{extra}.yaml
\end{lstlisting}
\begin{lstlisting}[caption={Bounding mesh type}]
gmsh.model.mesh.setSize:
- ipts: ...
gmsh.option.setNumber:
- Mesh.Algorithm: ...
- Mesh.Algorithm3D: ...
- General.NumThreads: ...
- Mesh.MeshSizeExtendFromBoundary: ...
- Mesh.MeshSizeFromPoints: ...
- Mesh.MeshSizeFromCurvature: ...
info: meta/(i|n|s){\d{2}}o_{extra}.yaml
\end{lstlisting}
\endgroup

{\color{gray}
\section{Hexahedral Teaser}

To come up.
}

\section{Conclusion}

HEXME is twofold.
On one hand, it is a tetrahedral dataset with feature entities.
On the other hand, it is a transparent workflow.
The main objective of HEXME is to provide relevant tetrahedral meshes for the fair assessment of hexahedral meshers, and associated auxiliary tools such as 3D frame fields.
The consistency of available meshes will likely evolve along with the progress of hexahedral techniques.
It is then crucial to expose the pipeline defining the mesh generation, such that HEXME dataset does not become obsolete.

The selected 63 CAD models come from several databases.
Their origin and license are recorded within a metadata file.
There are three categories of CAD models, and three types of meshes per CAD model.
The 189 meshes are produced thanks to a workflow that is defined with Snakemake.
The CAD features are reproduced by Gmsh as lower dimensional elements.
The meshes are expressed as vtk Datafile Version 2.0, in ASCII mode.

There are two ways to access HEXME tetrahedral meshes: either by downloading all of them in a 1.5GB \texttt{.zip} file, either by picking some of them from the catalog.
The \texttt{.zip} file also contains the log files from the mesh generation.
In addition to the meshes and log files, the catalog yields the metadata related to the CAD model, a summary about the mesh, and information related to the workflow.
The files that are involved in the workflow, are available on a GitHub repository.
From this git repository, it is possible to raise issue or/and pull-request, in order to improve the dataset, or the workflow.

In the future, the meshes should be tagged, like a release.
Besides, a \texttt{doi} should be set with \href{https://zenodo.org/}{Zenodo} whenever a release occurs.
This identification is crucial in order to keep track of the assessment of hexahedral methods.

\section*{Acknowledgements}

HEXME dataset is part of the ERC Starting Grant project "\emph{Algorithmic Hexahedral Mesh Generation}" -- ERC \href{https://www.algohex.eu/}{AlgoHex}, which is funded by the European Research Council under the European Union's Horizon 2020 research and innovation programme (Grant agreement No. 853343).

\bibliography{hexme}

\end{document}